\documentclass[%
 reprint,
 amsmath,amssymb,longbibliography,
 aps,
 prb,
]{revtex4-2}
\usepackage{hyperref}

\usepackage{graphicx}
\usepackage{amsmath}
\usepackage{rotating}
\usepackage{color}

\newcommand{\sruo}{Sr$_2$RuO$_4$}
\newcommand{\srrho}{Sr$_{2}$RhO$_{4}$}
\newcommand{\pros}{PrOs$_{4}$Sb$_{12}$}
\newcommand{\lsoc}{\lambda_{soc}}
\newcommand{\io}{\tilde \mu}

\newcommand{\down}{\downarrow}
\newcommand{\up}{\uparrow}
\newcommand{\spin}{\sigma}

\newcommand{\kv}{{\bf k}}
\newcommand{\qv}{{\bf q}}
\newcommand{\Qv}{{\bf Q}}
\begin{document}

\title{Theory of Strain-Induced Magnetic Order and Splitting of $T_c$ and $T_{\rm TRSB}$ in \sruo}

\date{January 2020}

\author{Astrid T. R\o mer$^1$, Andreas Kreisel$^2$, Marvin A. M\"{u}ller$^3$, P. J. Hirschfeld$^4$, Ilya M. Eremin$^3$, and Brian M. Andersen$^1$}
\affiliation{%
$^1$Niels Bohr Institute, University of Copenhagen, Lyngbyvej 2, DK-2100 Copenhagen,
Denmark
}%
\affiliation{$^2$Institut f\" ur Theoretische Physik, Universit\"at Leipzig, D-04103 Leipzig, Germany}
\affiliation{$^3$Institut f\"ur Theoretische Physik III, Ruhr-Universit\"at Bochum, D-44801 Bochum, Germany}
\affiliation{$^4$Department of Physics, University of Florida, Gainesville, Florida 32611, USA}

\date{\today}

\begin{abstract}

The internal structure of the superconducting state in \sruo ~remains elusive at present, and exhibits evidence for time-reversal symmetry breaking. Recent muon spin relaxation measurements under uniaxial strain have revealed an increasing splitting between the superconducting critical temperature $T_c$ and the onset of time-reversal symmetry breaking $T_{\rm TRSB}$ with applied strain [Grinenko {\it et al.}, \href{https://arxiv.org/abs/2001.08152}{ArXiv:2001.08152}]. In addition, static magnetic order is induced by the uniaxial strain beyond $\sim$1 GPa, indicating that unstrained \sruo ~is close to a magnetic quantum critical point. Here, we perform a theoretical study of the magnetic susceptibility and the associated pairing structure as a function of uniaxial strain. It is found that the recent muon relaxation data can be qualitatively explained from the perspective of spin-fluctuation mediated pairing and the associated strain-dependence of accidentally degenerate pair states in unstrained \sruo. In addition, while unstrained \sruo~features mainly $(2\pi/3,2\pi/3)$ magnetic fluctuations, uniaxial strain promotes $(\pi,\pm\pi/2)$ magnetism.

\end{abstract}
\maketitle

\section{Introduction}

Strontium Ruthenate, \sruo, has managed the remarkable feat
of remaining at the top of the superconductivity interest charts for more than two decades, albeit sometimes for the wrong reasons~\cite{Mackenziereview,Sigrist2005,Maenoreview,Kallin2012,Mackenzie2017}. Until recently, the material was a prime suspect in the search for topological chiral triplet superconductivity.  However, several new experimental results have challenged this picture~\cite{Pustogow19,ghosh2020thermodynamic,benhabib2020jump,grinenko2020split}, and the hunt is on to find a new consistent explanation for the panoply of observations on this fascinating material. These developments have not consigned \sruo~ to the junk heap of ``ordinary" unconventional superconductors, but led to a discussion of new ways that the system may be extraordinary, if not odd.

Prior to 2018-2019, the lack of an NMR Knight shift suppression upon entering the superconducting state~\cite{Ishida98_wrong}, in conjunction with the evidence for time-reversal symmetry breaking (TRSB) from $\mu$SR~\cite{Luke1998} and nonzero Kerr rotation measurements~\cite{Kapitulnik09}, pointed to chiral $p$-wave spin triplet superconductivity. However, there were several well-known “flies in the ointment” in the form of experimental evidence contradicting the $p$+i$p$-wave proposition. For example, NMR Knight shift results revealing constant susceptibility also in the case of out-of-plane magnetic fields~\cite{Murakawa2004}, the lack of chiral edge currents~\cite{Kirtley2007,Hicks2010,Curran2014}, the evidence for Pauli limiting critical fields~\cite{Yonezawa2013}, Josephson effects pointing to time-reversal symmetric superconductivity~\cite{Kashiwaya2019}, all did not straightforwardly support chiral $p$-wave as the preferred superconducting state in \sruo. In addition, several spectroscopic probes detected clear evidence for nodes in the superconducting gap, again not expected for a $p$+i$p$ phase~\cite{Firmo2013,Hassinger17,Kittaka2018}.

Recently, a reduction of the Knight shift for in-plane magnetic fields was discovered in the superconducting state of \sruo~\cite{Pustogow19,Ishida_correct}, a crucial result that is in contrast to earlier NMR investigations~\cite{Ishida98_wrong} due to control of 
sample heating effects. This drop in the spin susceptibility below $T_c$ is in stark contrast to the Knight shift expected for the chiral $p$+i$p$-wave phase, and suggests the realization of an even-parity condensate in this material. A standard single-component even-parity nodal superconducting order parameter is inconsistent with evidence for TRSB and conclusions of recent ultrasound measurements~\cite{ghosh2020thermodynamic,benhabib2020jump}. Both resonant ultrasound spectroscopy~\cite{ghosh2020thermodynamic} and ultrasound velocity measurements~\cite{benhabib2020jump} observe a discontinuity of the elastic constant $c_{66}$ at $T_c$, implying a two-component superconducting order parameter. With the odd-parity E$_u$ $p$-wave solution out of the running, attention has therefore turned to the even-parity (two-dimensional) E$_g$ representation or suitable combinations of one-dimensional representations~\cite{Roising2019,suh2019,kivelson2020proposal,Gingras18,Huang18,Huang19,Wang19,Kaba19,Acharya19,WangKallin20}. The E$_g$ state relies on interlayer pairing, a state hard to reconcile with the known weak interlayer coupling in \sruo. The latter possibility can be only relevant near degeneracy points, i.e. near regions of parameter space where two symmetry-distinct order parameters happen to be degenerate. An accidental two-component pairing state that seems currently consistent with the bulk of the experimental data, including the recent developments, is the $d+ig$ state\cite{kivelson2020proposal}.

However, at present only insight from additional experimental probes can help pinpoint the correct nature of the pairing state in \sruo. In this respect, recent muon spin relaxation ($\mu$SR) measurements under uniaxial strain are of significant interest~\cite{grinenko2020split}. Earlier, strain measurements had reported a notable absence of a cusp in $T_c$ vs. strain for both tensile and compressive strains, in disagreement with the chiral $p$+i$p$-wave scenario~\cite{Steppke17,Watson18}. More recently, measurements focusing on larger strains and the onset of the time-reversal symmetry-breaking $\mu$SR signal, have reported a clear strain-induced splitting between the superconducting $T_c$ and the onset temperature of the time-reversal symmetry-breaking phase, $T_{\rm TRSB}$~\cite{grinenko2020split}. In addition, uniaxial strain was shown to generate static magnetic order beyond $\sim$1 GPa~\cite{grinenko2020split}. 

Here, we focus on the evolution of the superconducting pairing instability as a function of uniaxial strain. We follow the theoretical approach of realistic spin-fluctuation-mediated pairing, including all bands near the Fermi level and sizable spin-orbit coupling (SOC) present in \sruo. This approach is known to mainly favor even-parity pair states (but also allows for odd-parity helical solutions) for interaction parameters  consistent with constraints from neutron scattering experiments\cite{Braden04,Steffens19}. Thus, importantly, it results in a near-degeneracy of symmetry-distinct pair states~\cite{RomerPRL,rmer2020fluctuationdriven}. This tendency for near-degeneracy is very distinct from, for example, iron-based systems and cuprates, favoring generally $s^{+-}$ and $d_{x^2-y^2}$, respectively~\cite{Hirschfeld_2011,Chubukov2012,Romer2015,RomerPRR2020}. Thus, \sruo~ is special in the sense that its Fermi surface features at least three competing nesting vectors that each prefer different superconducting pairing structures. Therefore the system is {\it pair-frustrated} with symmetry-distinct solutions lying close by in energy~\cite{RomerPRL,rmer2020fluctuationdriven}.  From this perspective, the accidental degeneracy scenario is particularly appealing.  

 \begin{figure}[t]
   	\includegraphics[angle=0,width=0.9\linewidth]{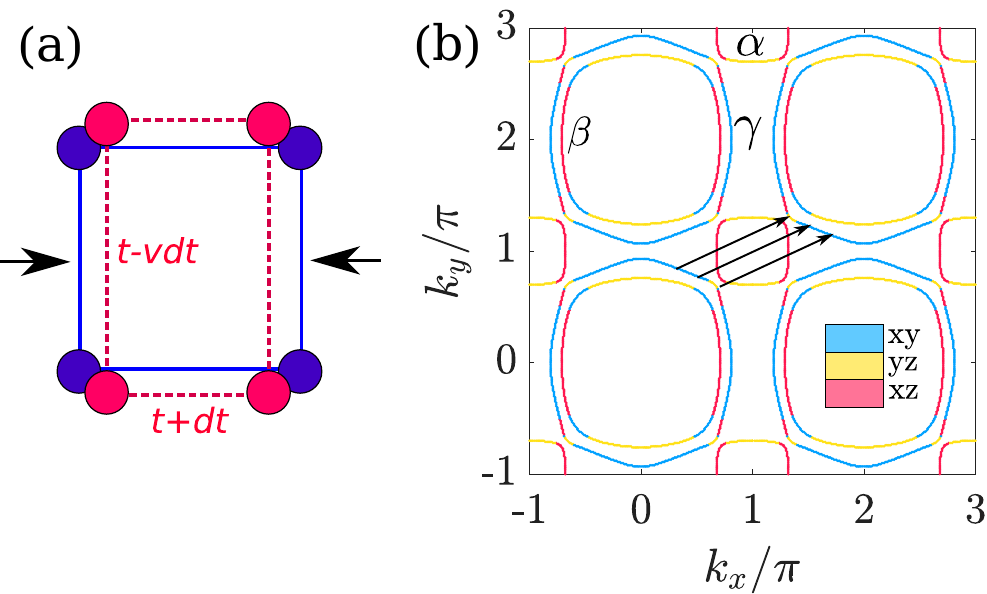}
\caption{(a) Sketch of applied strain $-p=\frac{dt}{t}$ (the minus sign indicates compression) as modeled by changes in the hopping constants along $(1 0 0)$ by $t +dt$ and along $(0 1 0)$ by $t -\textrm{v}dt$, where v is the in-plane Poisson ratio. (b)
The Fermi surface depicted in the extended zone scheme for $p=-9$ \%.  The common nomenclature of the Fermi pockets $\alpha,\beta,$ and $\gamma$ is indicated. The nesting vector $\Qv_3 \simeq(\pi,\pi/2)$ is indicated by three black arrows. Colors in panel (b) indicate majority orbital character, pink (xz), yellow (yz) and blue (xy).}
\label{fig:FS}
 \end{figure}

\section{Model and Method}

From Angular Resolved Photoemission Spectroscopy (ARPES) experiments~\cite{Tamai2019,Veenstra2014,Haverkort08,Zabolotnyy13} the normal state Fermi surface of \sruo~ constructed from the three Ru orbitals $d_{xz},d_{yz}$ and $d_{xy}$ is very well-established. In addition, the presence of a substantial SOC has been demonstrated. In this work, SOC is parametrized by $H_{SOC}=\lsoc \bf{L}\cdot\bf{S}$. Since time-reversal symmetry is preserved in the normal state, all energies are doubly degenerate and we can write the 
non-interacting Hamiltonian in block-diagonal form $\hat{H}=\sum_{{\bf k},\sigma} \Psi^\dagger(\kv,\sigma) (H_0+H_{SOC}) \Psi(\kv,\sigma)$, where each block is labeled by a pseudospin index $\spin=+(-)$ and describes one member of the Kramer's doublet. The matrices $H_0$ and $H_{SOC}$ are given by
\begin{eqnarray}
 H_0&=&\left( \begin{array}{ccc}
  \xi_{xz}(\kv) & g(\kv) &0 \\
  g(\kv) & \xi_{yz}(\kv) & 0 \\
  0 & 0 &   \xi_{xy}(\kv)
 \end{array}\right),
\end{eqnarray}
\begin{eqnarray}
 H_{SOC}&=&\frac{1}{2}\left( \begin{array}{ccc}
  0 & -i\spin\lsoc & i\lsoc \\
  i\spin\lsoc & 0 & -\spin\lsoc \\
  -i\lsoc & -\spin\lsoc &  0\\
 \end{array}\right),
 \label{eq:H0Hsoc}
\end{eqnarray}
within the basis
$\Psi(\kv,+)=[c_{xz\up}(\kv),c_{yz,\up}(\kv),c_{xy,\down}(\kv)]$, and 
$\Psi(\kv,-)=[c_{xz,\down}(\kv),c_{yz,\down}(\kv),c_{xy,\up}(\kv)]$.
Here, $c_{\mu,s}(\kv)/c^\dagger_{\mu,s}(\kv)$ are electronic annihilation/creation operators of orbital character $\mu$ and spin $s$. The pseudospin is $\spin=+(-)$ for the up (down) block Hamiltonian stated in Eq.~(\ref{eq:H0Hsoc}).
The electronic dispersions are given by
$\xi_{xz}(\kv)=-2t_1\cos k_x -2t_2\cos k_y -\mu$, 
$\xi_{yz}(\kv)=-2t_2\cos k_x -2t_1\cos k_y -\mu$,
$\xi_{xy}(\kv)=-2t_3(\cos k_x +\cos k_y) -4t_4\cos k_x \cos k_y-2t_5(\cos 2k_x +\cos 2k_y) -\mu$,
 with the hopping constants $\{t_1,t_2,t_3,t_4,t_5\}=\{88,9,80,40,5\}$ meV\cite{Cobo16,Zabolotnyy13}. Orbital hybridization between $xz$ and $yz$ is parametrized by $t'$ in  $g(\kv)=-4t'\sin(k_x)\sin(k_y)$ which is set to $t'=4.4$ meV($=0.05t_1$)\cite{WangKallin20}. Spin-orbit coupling is set to $\lsoc=35$ meV and a lower value of 10 meV for comparison. Note that due to the use of renormalized hopping constants, $\lsoc=35$ meV corresponds to $0.4t_1$.
 Our model is restricted to two dimensions and we quantify compressive in-plane strain by a percent-wise change in the hopping parameters as sketched in Fig.~\ref{fig:FS}(a). Along the $(1 0 0)$ direction, the hopping parameters undergo a relative increase of $\frac{dt}{t}=-p>0$, while a relative decrease of $\textrm{v}p$ is implemented along the $(0 1 0)$ direction. Here $\textrm{v}=0.51$ refers to the low-temperature in-plane Poisson ratio recently reported by Barber {\it et al.}\cite{Barber2019}. The average charge density is kept constant at four electrons at all strains by adjusting the chemical potential $\mu$.

We investigate the influence of strain on spin-fluctuation mediated superconductivity. The effective electron-electron interaction in the Cooper channel from the multi-orbital Hubbard Hamiltonian due to spin fluctuations was derived in Ref.~\onlinecite{RomerPRL}. It includes intra- and inter-orbital Coulomb interactions and Hund's coupling terms and effective interactions mediated by spin-fluctuations in the multi-orbital random-phase approximation
\begin{equation}
 \hat{H}_{int}=\frac{1}{2}\!\sum_{ \kv,\kv' \{\tilde \mu\}}\!\!\Big[V(\kv,\kv')\Big]^{\io_1 , \io_2 }_{\io_3,\io_4 }  c_{\kv \io_1 }^\dagger  c_{-\kv \io_3 }^\dagger c_{-\kv' \io_2 } c_{\kv' \io_4 },
 \label{eq:Heff}
\end{equation}
with the pairing interaction given by
\begin{eqnarray}
\Big[V(\kv,\kv')\Big]^{\io_1 , \io_2 }_{\io_3,\io_4 } &=&\Big[U\Big]^{\io_1 , \io_2 }_{\io_3,\io_4 }+\Big[U\frac{1}{1-\chi_0U}\chi_0U\Big]^{\io_1 \io_2}_{\io_3 \io_4}(\kv+\kv') \nonumber \\
&& -\Big[U\frac{1}{1-\chi_0U}\chi_0U  \Big]^{\io_1\io_4}_{\io_3 \io_2}(\kv-\kv') .
\label{eq:Veff}
\end{eqnarray}
The label $\io \:= (\mu,s)$ is a joint index for orbital and electronic spin, and contributions from bubble and ladder diagrams are accounted for.  In Eq.~(\ref{eq:Veff}), $\chi_0$ refers to the real part of the generalized susceptibility 
\begin{eqnarray}
&& [\chi_0]^{\tilde \mu_1,\tilde \mu_2}_{\io_3,\io_4}(\qv,i\omega_n)=
\frac{1}{N}\int_0^\gamma d \tau e^{i \omega_n \tau}\nonumber \\ &&
\quad\sum_{\kv,\kv'} 
 \langle T_\tau c^\dagger_{\kv- \qv \mu_1s_1} (\tau)
c_{\kv \mu_2s_2}  (\tau) c^\dagger_{\kv'+ \qv \mu_3s_3}  c_{\kv' \mu_4s_4} \rangle_0,
 \end{eqnarray}
which is evaluated at zero energy and includes the effects of SOC and strain. Leading and sub-leading superconducting instabilities at the Fermi surface 
are determined by a projection of the interaction Hamiltonian Eq.~(\ref{eq:Heff}) to band- and pseudospin-space, followed solving the (linearized) BCS gap equation
\begin{eqnarray}
  -\int_{FS} d \kv_f^\prime \frac{1}{v(\kv_f^\prime)} \Gamma_{l,l'}(\kv_f,\kv_f^\prime)\Delta_{l'}(\kv_f^\prime)=\lambda \Delta_l(\kv_f)
  \label{eq:LGE}
\end{eqnarray}
for the eigenvalue $\lambda$ and the gap function $\Delta_{l}(\kv_f)$ at wave vectors 
 $\kv_f$ on the Fermi surface. The Fermi surface is discretized by approximately 1000 wave vectors and $v(\kv_f)$ is the Fermi speed.
The pairing kernel in band space is given by $\Gamma_{l,l'}(\kv_f,\kv_f^\prime)$, with the spin information carried by the subscripts $l,l'=0,x,y,z$ which refers to the ${\bf d({\bf k})}$-vector~\cite{SigristUeda} in {\it pseudospin} space. 
For further details we refer to Ref.~\onlinecite{RomerPRL}.
As a consequence of Pauli's exclusion principle, pseudospin singlet states ($l=0$) are even in parity, while pseudospin triplet solutions ($l=x,y,z$) are odd-parity states. Characterization in terms of irreducible representations of the $D_{4h}$ group breaks down as a result of lowering of the point group symmetry to $D_{2h}$ because of the applied strain, see also Fig.3(b). Even-parity solutions are characterized by $A_{1g}$ ($B_{1g}$) of $D_{2h}$ for solutions without (with) nodes along the axes $k_x=0$ and $k_y=0$. For simplicity, we denote all odd-parity solutions by $B_u$. This encompasses solutions of the form  $k_y {\bf \hat x}+k_x {\bf \hat y}$,$k_x {\bf \hat x}+k_y {\bf \hat y}$, and $k_y {\bf \hat z}$ with a complicated momentum structure with formation of higher order nodes. In the $D_{2h}$ point group there are no symmetry-protected two-component solutions, thus the occurrence of degenerate solutions arises only accidentally.
When referring to an irreducible representation of the $D_{4h}$ group, we will use the notation $s'$ ($A_{1g}$ in $D_{4h}$),  $g$ ($A_{2g}$ in $D_{4h}$),  $d_{x^2-y^2}$ ($B_{1g}$ in $D_{4h}$),  $d_{xy}$ ($B_{2g}$ in $D_{4h}$) and {\it helical} for one-component odd-parity solutions of $D_{4h}$ and {\it chiral} for the two-component $E_u$ solution in $D_{4h}$. The labels $A_{1g}$, $B_{1g}$ and $B_{u}$ henceforth refer to an irreducible representation of the $D_{2h}$ group.

\section{Results}
 
\begin{figure}[t]
 \centering
   	\includegraphics[angle=0,width=\linewidth]{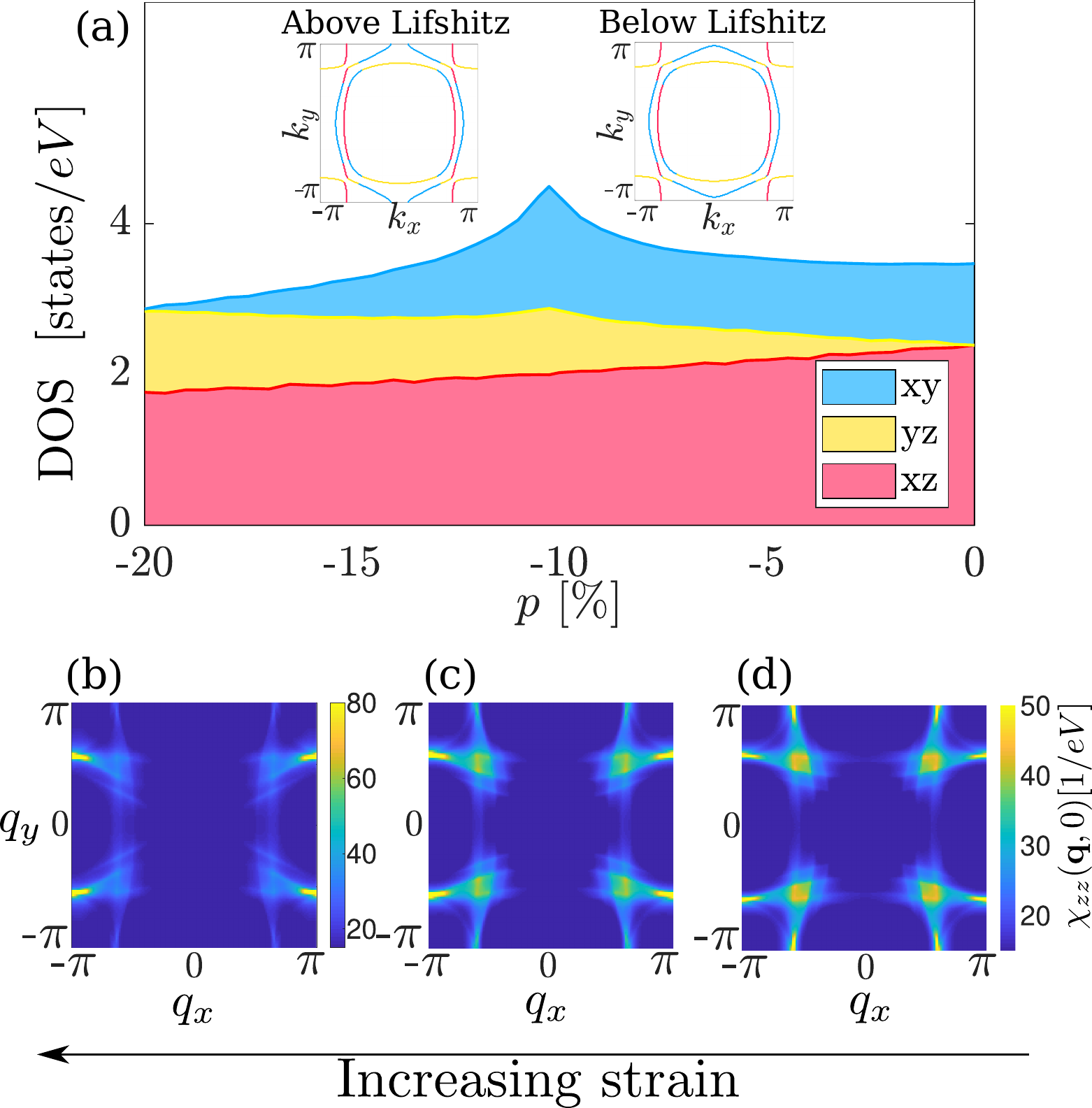}
\caption{(a) Density of states at the Fermi level for the three orbitals $xy,yz,xz$ as a function of applied strain. Fermi surfaces below ($p=-9$ \%) and above ($p=-12$ \%) the Lifshitz transition are depicted by insets, with the dominating orbital character at the Fermi surface indicated by colors.
(b-d) Real part of the spin susceptibility $\chi_{ zz}(\qv,0)$ for applied strain (b) $p=-12$ \%, (c) $p=-6$ \% and (d) $p=-0.1$ \% for the intermediate coupling parameters $U=140$ meV and $J/U=0.16$. Note the difference in color scale for (b). 
}
\label{fig:susc}
\end{figure}

First, we address the effect of strain on the normal state. The change in the density of states at the Fermi level (DOS) as a function of strain $p$ is shown in Fig.~\ref{fig:susc}(a). For a band with $\lsoc=35$ meV, the $\gamma$-pocket touches the van Hove points at $(0,\pm \pi)$ at the strain value $p_c=-10.5$ \%. At this strain value, the DOS of the $xy$ orbital peaks and the normal state Fermi surface undergoes a Lifshitz transition, where the $\gamma$-pocket splits at the van Hove points. This is illustrated in the insets of Fig.~\ref{fig:susc}(a) and is in agreement with recent ARPES measurements~\cite{Sunko2019} as well as {\it ab initio} calculations~\cite{Hsu16,Steppke17,Barber2019}. In Fig.~\ref{fig:FS}(b) the Fermi nesting vector $\Qv_3=(\pi,\pi/2)$ is indicated~\cite{Cobo16,RomerPRL}.
In unstrained samples, this nesting is only subdominant to the main peak at 
$\Qv_1=(2\pi/3,2\pi/3)$, but time-of-flight neutron scattering has observed magnetic structures at $\Qv_3$~\cite{Iida11}. 
Here we argue that the effect of tuning the $xy$ orbital through van Hove points causes the $\Qv_3$ peak to dominate the spin-fluctuation spectrum, see Fig.~\ref{fig:susc}. 
This occurs because Fermi surface nesting builds up as a consequence of the deformation of the $\gamma$-pocket and becomes the dominant contribution in the spin response, as shown in Fig.~\ref{fig:susc}(b-d) for the out-of-plane ($zz$) spin component for increasing strain.
The SDW order associated with the $\Qv_3\simeq(\pm \pi,\pm\frac{\pi}{2})$ ordering vector is approximately described by period two along the $(100)$ direction and period four along the $(010)$.

 \begin{figure*}[t]
 \centering
   	\includegraphics[angle=0,width=\linewidth]{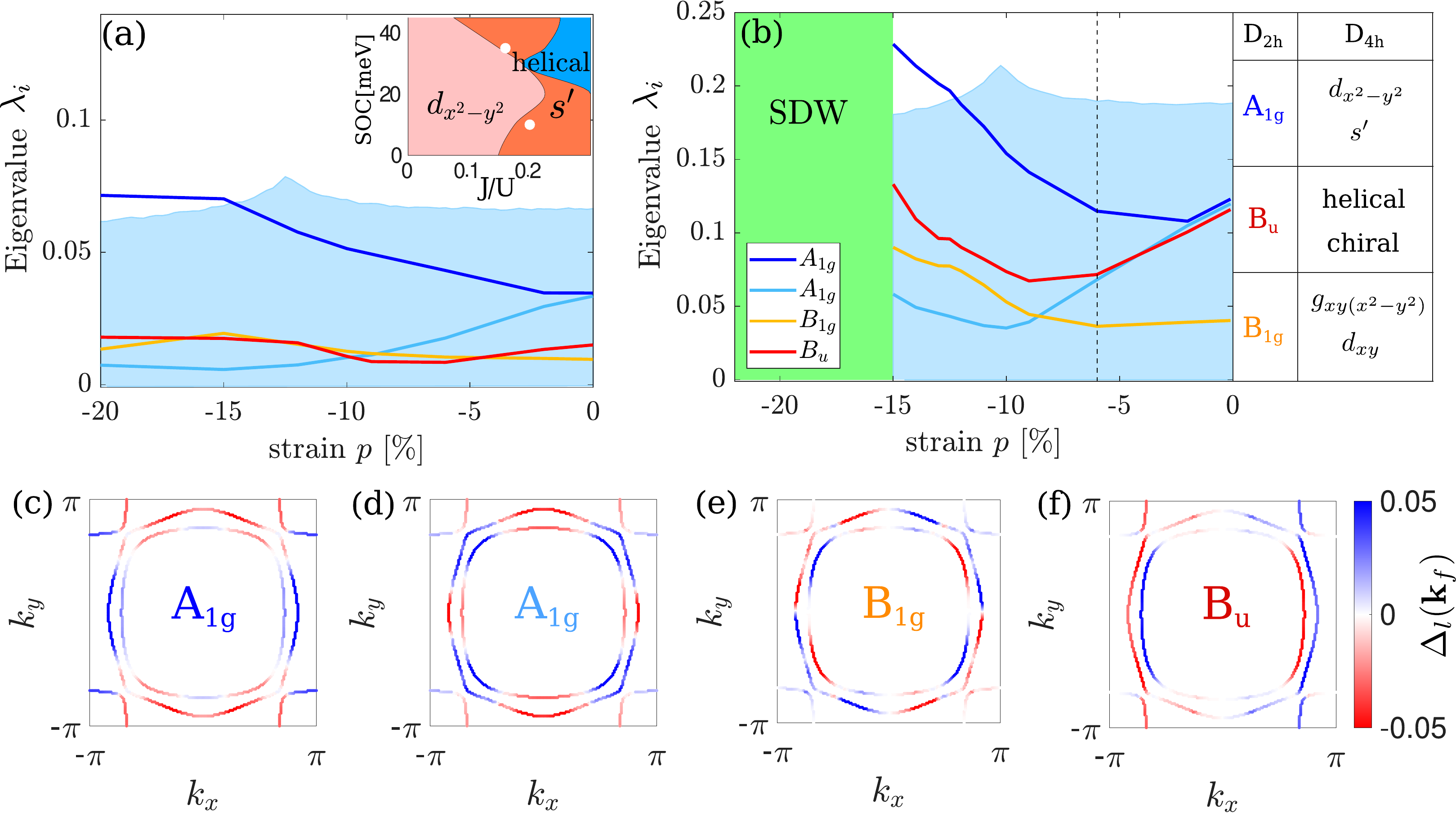}
\caption{(a,b) Leading eigenvalues of the linearized gap equation as a function of strain for (a) $\lsoc=10$ meV, $U=100$ meV and $J/U=0.2$ and (b) $\lsoc=35$ meV, $U=140$ meV and $J/U=0.16$. 
The two leading solutions of $A_{1g}$ are shown by dark and light blue, $B_{1g}$ in orange and the leading odd parity channel is shown in red. The connection to the irreducible representation of the $D_{4h}$ group is shown in the table to the right of (b). Other subleading solutions are not shown.  The  blue background color displays the DOS at the Fermi level of all three orbitals (arbitrary scale). The inset of (a) displays the phase diagram as a function of $J/U$ and SOC in zero strain (with $t'=0$). The two white dots indicate the (unstrained) starting point of the strain calculation in panels (a) and (b), reproduced from Ref.~\onlinecite{RomerPRL}. (c-f) the gap solutions at the Fermi level at strain $p=-6$ \%, indicated by the dashed line in panel (b). }
\label{fig:strong}
 \end{figure*}

The superconducting response to applied strain is shown in Fig.~\ref{fig:strong}
in the case of intermediate-coupling strengths of $U$ and $J$ with the superconducting pairing given by spin fluctuations as expressed in Eqs.~(\ref{eq:Heff}-\ref{eq:Veff}). In Fig.~\ref{fig:strong} we display results for two different values of $\lsoc$ of $10$ meV and $35$ meV, and for simplicity plot only the leading solution in each channel with the exception of the $A_{1g}$ channel, where the two leading solutions are shown. In both the case of weak and strong SOC, strain gives rise to a splitting between the single leading and all the subleading superconducting instabilities. 
The leading instability appears in the $A_{1g}$ channel. In the unstrained case, this solution corresponds to the $d_{x^2-y^2}$ state, which is nearly degenerate with the nodal $s'$-wave solution~\cite{RomerPRL}, as displayed in the inset of Fig.~\ref{fig:strong}(a). The eigenvalue of this solution increases upon applied strain; for $\lsoc=10$ meV the increase is immediate, while for $\lsoc=35$ meV  it occurs for strain values $|p|>2$ \%. 

By contrast, all the subleading superconducting solutions are rather inert to strain, or pushed down for a large range of applied strains. While the leading solution continues to display $A_{1g}$ symmetry for all strains, the order of subleading solutions may change as a function of strain. In fact, an overall preference for odd-parity solutions upon strain occurs; the second-leading solution is of $B_{u}$ type in the case of strong SOC for $|p|>6$ \%, see Fig.~\ref{fig:strong}(b). For smaller values of SOC, the odd-parity solutions remain in close competition with other subleading solutions, as shown in Fig.~\ref{fig:strong}(a). In particular, the $B_{1g}$ solutions, which also have nodes along the zone axes appear close in energy to the odd-parity solution.

In Fig.~\ref{fig:strong}(c,d) we show the two leading $A_{1g}$ solutions in the case of $\lsoc=35$ meV and $|p|=6$ \%. These gap structures are connected to the $d_{x^2-y^2}$ and nodal $s'$-wave solutions of the unstrained system, respectively. The $B_{1g}$ solution at $|p|=6$ \% shown in Fig.~\ref{fig:strong}(e) is an even-parity solution with nodes along the zone axes, and resembles the $g_{(x^2-y^2)xy}$ state of the zero strain case.
The $\bf \hat y$-component of the $B_u$ leading odd-parity solution is shown in Fig.~\ref{fig:strong}(f). Surprisingly, we find the largest gap magnitudes on Fermi surface segments along the $k_x=0$ axis with no visible advantage of the large DOS region close to $(0,\pm\pi)$. This is different in the weak-coupling regime, as discussed below.

The recent experimental $\mu$SR results of Ref.~\onlinecite{grinenko2020split} find a splitting between $T_c$ and $T_{\rm TRSB}$ upon strain. Our results of Fig.~\ref{fig:strong} provide a possible explanation for this experimental finding due to the observed splitting between leading and subleading superconducting instabilities as a function of applied strain. The exact structure of the leading and subleading solutions, as well as their quantitative splitting, depends on the band structure, SOC and $U/J$ coupling strengths. However, a clear property of the theory is the "splitting-off" of the leading instability from all the subleading states which are less affected by the strain. Therefore, we expect quite generally that $T_c$ is enhanced by strain while $T_{\rm TRSB}$ remains rather unaffected by the strain field. Within the current scenario, any TRSB signal must arise from pinned supercurrents near defects and other lattice imperfections because the states considered are not chiral.

Recently, a proposal of an accidental degeneracy between $d_{x^2-y^2}$ and $g$-wave was presented~\cite{kivelson2020proposal} as a possibility to reconcile a number of experimental observations in unstrained \sruo, notably the $c_{66}$ jump in the $B_{2g}$ shear modulus in ultrasound~\cite{ghosh2020thermodynamic,benhabib2020jump} and the concurrent absence of a jump in the $B_{1g}$ channel. In general, spin-fluctuation pairing based on onsite interactions $U$ and $J$ find a $g$-wave solution which is strongly suppressed compared to the other even-parity solutions, see the orange curve of Fig.~\ref{fig:strong}(a,b). We point out that there are several other $A_{1g}$ and $B_u$ solutions above the $B_{1g}$ solution, which we have not shown in Fig.~\ref{fig:strong} for simplicity. While on  a square lattice with repulsive interactions it has been shown that significant nearest-neighbor Coulomb interaction favors $g$-wave pairing\cite{Raghu2012}, it remains to be determined what microscopic interactions can enhance this channel  in this multiband case. However, even in a case where $g$-wave pairing became subleading (and nearly degenerate) to $d_{x^2-y^2}$-wave in the unstrained case, based on the strain-dependence of the $d$- and $g$-wave solutions from Fig.~\ref{fig:strong}, we expect qualitatively similar strain-behavior with an enhanced leading $d$-wave solution, and a roughly strain-independent subleading $g$-wave solution. Finally, it remains very interesting to determine experimentally whether the tendency for strain-enhanced odd parity solutions, resulting in e.g. $\Delta_0+i(\Delta_x {\bf \hat x}+\Delta_y {\bf \hat y})$ superconductivity as in Fig.~\ref{fig:strong}, can be realized for sufficiently large strain.  

The green region in Fig.~\ref{fig:strong}(b) indicates the regime of  spin-density-wave (SDW) order.  In our formalism, the onset of a SDW instability is identified by a divergence of the spin susceptibilities, which, for fixed interaction strength, is triggered by the increase of applied strain. For the current band structure and interaction strengths of Fig.~\ref{fig:strong}(b), the SDW instability occurs at $|p_{\rm SDW}| > 15 $ \%. In Fig.~\ref{fig:strong}(a) the interaction strength $U$ is lower and hence there is no signature of an approaching SDW instability. A recent theoretical study of the superconducting pairing within the one-band Hubbard model found that the RPA formalism overestimates the instability channels as one gets very close to the magnetic instability~\cite{RomerPRR2020}. 
Therefore, we focus the pairing calculations on strain values below $|p|=15$ \%. 
A limitation of the current approach close to the SDW instability is the absence of competition between superconductivity and SDW order and low-energy fluctuations~\cite{romer2016}. In particular, since we do not calculate the pairing solutions self-consistently, any feedback effect of SDW is not taken into account. However, it is clear that the onset of SDW order would decrease the DOS at the Fermi level, and thereby suppress Cooper pairing. As a consequence, we expect $T_c$ to gradually decrease upon entering the SDW phase, consistent with experiments~\cite{grinenko2020split}. As mentioned above in the discussion of Fig.~\ref{fig:susc}(b-d), the predicted structure of the SDW order is of the approximate $(\pm \pi,\pm \pi/2)$ form.

 \begin{figure}[b]
 \centering
\includegraphics[angle=0,width=\linewidth]{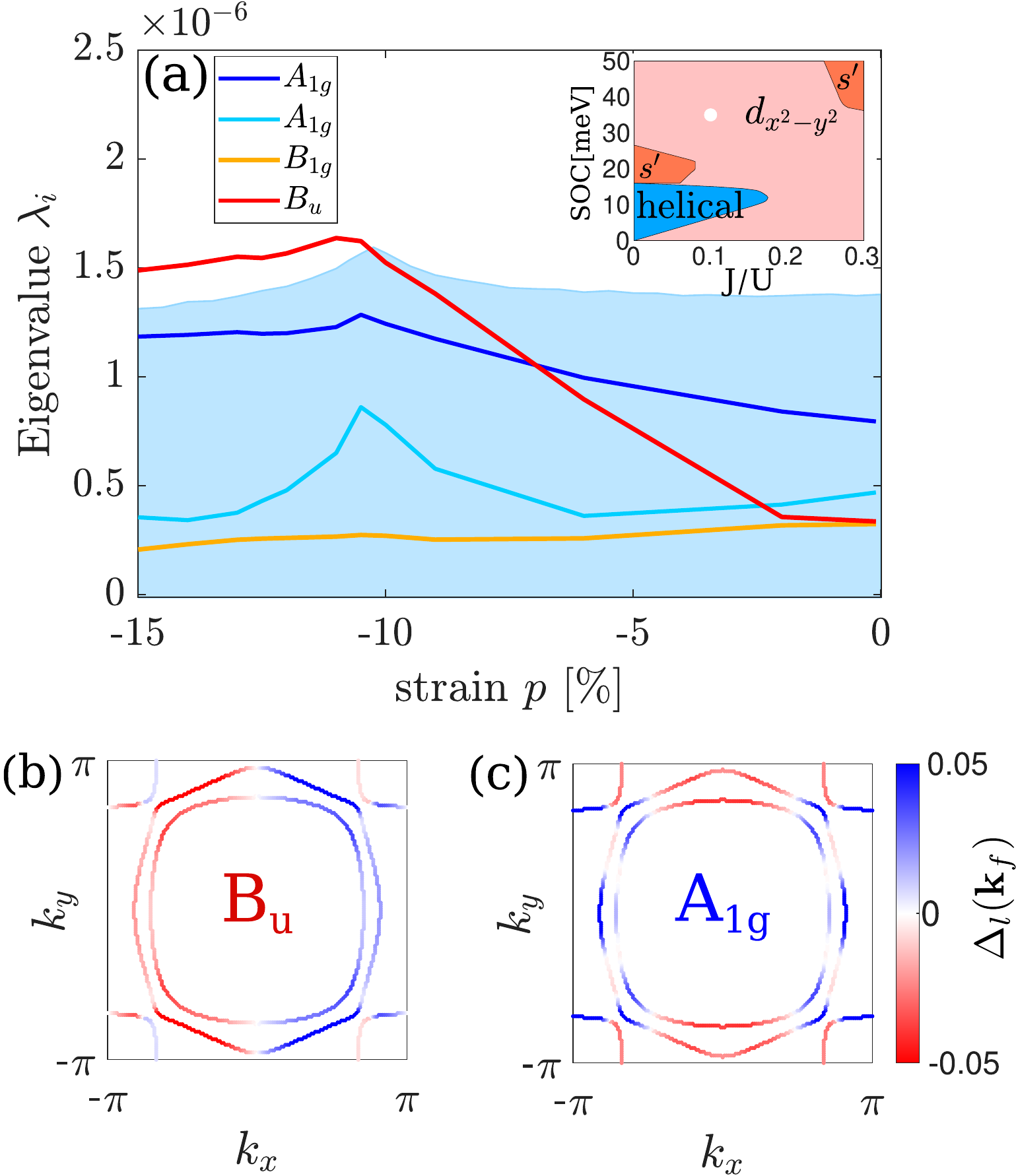}
\caption{(a) Leading eigenvalues of the linearized gap equation as a function of strain for $\lsoc=35$ meV in the weak-coupling limit of $U=1$ meV and $J/U=0.1$. The two leading $A_{1g}$ and the leading $B_{1g}$ and $B_u$ solutions are shown. Other subleading solutions are not shown.
The blue background displays the total DOS at the Fermi level (arbitrary scale). The inset displays the phase diagram as a function of $J/U$ and SOC in zero strain, reproduced from Ref.~\onlinecite{rmer2020fluctuationdriven}. The white dot marks the (unstrained) starting point of the strain calculation. (b,c) Momentum dependence of the leading odd ($B_u$)- and even-parity ($A_{1g}$) solution at $p=-9$\%.}
\label{fig:weak}
 \end{figure}

As evident from Fig.~\ref{fig:strong}, there is no direct correlation between the superconducting eigenvalues and the peak in the DOS. This is a feature of the intermediate-coupling regime owing to important changes in the spin-fluctuation spectrum as a function of strain, as visualized by the $z$-component of the susceptibility in Fig.~\ref{fig:susc}(b-d). When changes in the spin-fluctuation mediated pairing dominate the density of states effect, the leading and subleading instabilities are split. Our findings show that only one instability provides a favorable match to the enhanced nesting structures. With the complicated orbital- and spin-structure of the pairing kernel, Eq.~(\ref{eq:Veff}), it is not  {\it a priori} obvious which of the solutions will be favored under strain, but we find this behavior to be generic in the limit of intermediate-coupling strengths.


This conclusion is different from the case of weak couplings~\cite{Steppke17,LiuWang17,Hsu16}. In Fig.~\ref{fig:weak} we display a typical example of the superconducting response to applied strain in the limit of weak interactions ($U=1$ meV and $J/U=0.1$). As seen, it is dominated by the increase in the DOS at the Fermi level. The $A_{1g}$ and $B_u$ solutions increase initially, followed by a smaller decrease above the  Lifshitz transition point, whereas the $B_{1g}$ solutions remain largely unaffected by strain. As in Fig.~\ref{fig:strong}, we plot only the leading solution in each channel with the exception of the $A_{1g}$ channel, where the two leading solutions are shown. In the vicinity of the Lifshitz transition, the three leading solutions are in fact all odd-parity order parameters. These are nearly degenerate and all take advantage of the nesting vector $\Qv_3=(\pi,\pi/2)$ depicted in Fig.~\ref{fig:FS}(b). 
This can be seen from Fig.~\ref{fig:weak}(b), which depicts the leading solution in the $\bf \hat y$ channel. The Fermi surface regions with the largest gap values match the regions connected by the nesting vector $\Qv_3$. 
Since the nesting vector connects symmetry-related momentum states on the same band, it does not promote even-parity states. This is visualized by the subleading even-parity $A_{1g}$ solution in Fig.~\ref{fig:weak}(c) which has a more elaborate nodal structure and does not provide an obvious match to the $\Qv_3=(\pi,\pi/2)$ nesting. Since the most prominent vector in the strained case is $\Qv_3$, it is reasonable that the odd-parity solutions become favored in this limit. 
We conclude that in the weak-coupling limit, 
the favorisation of odd-parity solutions is a result of the new prominent nesting in the strained case and
the signature of the DOS increase is visible in the superconducting channel. This behavior agrees roughly with the weak-coupling calculations in Ref.~\onlinecite{Steppke17}.
The simple picture breaks down in the intermediate-coupling regime discussed above, because of further orbital details in the pairing structure which are not captured by an increase of one single nesting vector. 
\section{Conclusions}

We have calculated the evolution of the hierarchy of superconducting solutions within spin-fluctuation mediated pairing under uniaxial strain. In the intermediate-coupling regime, it is found that the leading instability splits-off from all the subleading solutions as a function of strain, in qualitative agreement with recent muon spin relaxation measurements. This is in contrast to the weak-coupling regime where all pairing solutions follow the density of states evolution with strain. The proposed scenario for superconductivity in \sruo~relies on the material being close to an accidental degeneracy of pair states in the unstrained case, and naturally explains the emergence of static magnetic order with sufficiently large uniaxial strain.      

\section{Acknowledgements}

We acknowledge useful discussions with S. Mukherjee and S. H. Simon. A.T.R. and B.M.A. acknowledge support from the Carlsberg Foundation.
P. J. H. was
supported by the U.S. Department of Energy under Grant
No. DE-FG02-05ER46236.
 \bibliography{bibliography_strain}
\end{document}